\documentclass{article}
\usepackage[english]{babel}
\usepackage{arxiv}

\usepackage[utf8]{inputenc} 
\usepackage[T1]{fontenc}    
\usepackage{hyperref}       
\usepackage{url}            
\usepackage{doi}            
\usepackage{amsfonts}       
\usepackage{nicefrac}       
\usepackage{microtype}      

\usepackage{amsmath, amssymb, graphicx, algorithm2e}
\usepackage[round]{natbib} 
\usepackage{longtable}
\usepackage{booktabs}
\usepackage{siunitx}
\usepackage{algorithmic}
\usepackage{textcomp}
\usepackage{xcolor}
\usepackage{tikz}
\usetikzlibrary{shapes.geometric}
\usepackage{listings}
\usepackage{subcaption}
\usepackage{pdflscape}
\usepackage{afterpage}
\usepackage{array}
\usepackage{inconsolata}
\usepackage{tabularx}
\usepackage{caption}

\title{Evaluating LLM Generated Detection Rules in Cybersecurity}

\author{
    Anna Bertiger,
    Bobby Filar,
    Aryan Luthra,
    Stefano Meschiari,
    Aiden Mitchell,
    Sam Scholten,
    Vivek Sharath
    \\ \\
	\textbf{Sublime Security}\\
	Washington, DC, USA\\
	\texttt{\{anna.b, bobby, aryan, stefano, aiden, sam, vivek\}@sublimesecurity.com}
}

\definecolor{codegreen}{rgb}{0,0.6,0}
\definecolor{codegray}{rgb}{0.5,0.5,0.5}
\definecolor{codepurple}{rgb}{0.58,0,0.82}
\definecolor{backcolour}{rgb}{0.95,0.95,0.92}

\lstdefinelanguage{MQL}{
  keywords={type, disposition, sender,
  body, file, any, all, of, from, where, and, or, not, map, filter,
            length, any, all, flatten, to_string, lower, matches, starts_with, ends_with, contains},
  keywordstyle=\color{blue}\bfseries,
  comment=[l]{//}, 
  commentstyle=\color{codegreen},
  stringstyle=\color{codepurple},
  sensitive=false, 
  morecomment=[s]{/*}{*/}, 
  morestring=[b]", 
}

\begin{document}
\maketitle

\begin{abstract}
LLMs are increasingly pervasive in the security environment, with limited measures of their effectiveness, which limits trust and usefulness to security practitioners. Here, we present an open-source evaluation framework and benchmark metrics for evaluating LLM-generated cybersecurity rules. 

The benchmark employs a holdout set-based methodology to measure the effectiveness of LLM-generated security rules in comparison to a human-generated corpus of rules. It provides three key metrics inspired by the way experts evaluate security rules, offering a realistic, multifaceted evaluation of the effectiveness of an LLM-based security rule generator. 

This methodology is illustrated using rules from Sublime Security's detection team and those written by Sublime Security's Automated Detection Engineer (AD\'E), with a thorough analysis of AD\'E's skills presented in the results section.\end{abstract}

\keywords{evaluating LLMs \and agents \and security rules}

\section{Introduction}
AI agents are popular, and vendors are latching on the to the idea that idea that they will revolutionize the detection experience. But agents are not without their skeptics, with experienced detection engineers concerned that they produce “plausible sounding nonsense”. Like hiring a new employee, it’s important to measure whether the agent can do the job. Without these measurements, we can’t build trust in the agents’ work, and thus will not save any time or energy by using them. 

This paper describes a methodology for measuring the effectiveness of LLM generated security rules, and illustrates this method using Sublime Security's Automated Detection Engineer, AD\'E, which is an agentic system that writes queries in Message Query Language (MQL) to detect malicious emails. We evaluate the quality of queries by measuring detection accuracy of samples flagged by the query using a database of labeled email samples, and measuring the robustness of queries written. In addition, we measure the cost of operating the system to optimize cost vs. benefit. 

We also compare the quality of queries written by AD\'E with existing human-developed detection rules that experts at Sublime have published and maintained to help protect users \citep{corefeed}. We choose a selection of rules, and, one at a time, remove a rule from AD\'E's knowledge base, and ask AD\'E to construct a detection for a true positive (TP) based on that rule. 

\section{Background and Related Work}
Measuring the quality of LLM output is notoriously difficult and context dependent; see, for example, the survey of Chang et al. \citet{chang2024survey}, and the recent paper on measuring LLM based systems from Rudd, Andrews and Tully \citet{rudd2025practicalguideevaluatingllms}. Measuring the quality of automatically generated security detection rules is no exception. 

We draw inspiration from two main sources: the literature that measures the quality of LLM-generated code and the literature that measures the quality of cybersecurity detections. In addition, we take some inspiration from supervised machine learning, by holding out human-created rules from AD\'E's knowledge base in order to compare AD\'E to a human expert. Unlike with supervised learning, this is not entirely a fair comparison, as the human-generated rules are designed by humans fitting a broader collection of data and total background domain knowledge than AD\'E is given. 

The code quality literature has a number of papers focused on creating robust test sets where code quality is measured by the ability to change existing tests from failing to passing starting with existing code, test cases and issue descriptions, for example \citep{jimenez2024swebench} and \citep{zhang2025swebenchgoeslive}. These measurements produce leader boards where LLM-based coding agents compete to produce the best code. Chen et al. \citet{chen2024largelanguagemodelsconfront} examine LLM performance in repository-level program repair tasks, revealing significant gaps between isolated code generation and real-world debugging scenarios that require understanding of a broader codebase.

Recently, we have also seen the development of code robustness measurements.  Li et al. \citet{li2025enhancingrobustnessllmgeneratedcode} develop a measurement of the robustness of LLM-generated code based on counting conditional statements, which are assumed to increase robustness by checking for edge cases and try/except statements, which are assumed to indicate a lack of robustness if they are needed.

\subsection{Evaluating LLMs in Code Generation Tasks}
Research into quantifying the output of code-generation LLMs has grown rapidly from measuring the similarity of generated code to that of an oracle to capture functional correctness. The work of Chen et al.\citet{chen2021evaluatinglargelanguagemodels} introduced the HumanEval benchmark, along with the $pass@k$ metric, which measures the probability that one of the K-generated samples passes a series of unit tests.

The concept of functional correctness has become foundational in the evaluation of code generation models. However, researchers continue to expand this metric to better reflect the real world concerns of introducing LLM-generated code into production environments \citep{jimenez2024swebench}\citep{zhang2025swebenchgoeslive}.

\subsubsection{Secure Code Generation}
Currently, there is a growing body of work that measures the security of LLM-generated code. Benchmarks such as CWEval\citep{peng2025cweval} and CodeGuard+\citep{fu2024constrained} highlight the importance of capturing both functional correctness and security of code generation tasks. This is accomplished using an expanded $pass@k$ metric that penalizes solutions that are secure but functionally incorrect (e.g., a no-op solution). Most recently, Dilgren et al. \citet{dilgren2025secrepobench} presented SecRepoBench, which evaluates LLMs on secure code generation within real-world repository contexts, highlighting the importance of testing generated code in production-like environments.

\subsubsection{Code Robustness}
Moving beyond the generation of secure code, research such as the RobGen\citep{li2025enhancingrobustnessllmgeneratedcode} framework introduces metrics and techniques to evaluate the generated code's ability to handle edge cases or incorrect input, capturing the code's robustness to unexpected scenarios. Likewise, earlier work by Wang et al. \citet{wang2022recode} introduced ReCode, which evaluates the robustness of code generation models through perturbation-based testing, establishing foundational methods for assessing the reliability of generated code.

Similarly, Pasini et al. \citet{pasini2024evaluatingimprovingrobustnesssecurity} demonstrate that LLM-generated code often exhibits brittleness that can be evaluated and improved through iterative refinement, strengthening our approach to measuring the robustness of detection rules.

\subsection{LLMs for Cybersecurity Tasks}
The use of LLMs in the cybersecurity space has also experienced rapid growth, with applications in both offensive and defensive tasks.

For offensive applications, Dawson et al. \citet{dawson2025airtbenchmeasuringautonomousai} introduce AIRTBench, which measures autonomous AI red teaming capabilities in language models, evaluating their ability to identify and exploit vulnerabilities. This work highlights the dual-use nature of LLMs in security, where the same capabilities that enable the generation of defensive rules can also be used for adversarial testing. Fang et al. \citet{fang2024llmagentsautonomouslyhack} demonstrate that LLM agents can autonomously hack websites, discovering and exploiting vulnerabilities without human intervention. These offensive capabilities underscore the arms race nature of cybersecurity, where defenders must anticipate LLM-powered attacks when designing detection rules.

On the defensive side, key applications include threat intelligence, vulnerability and malware analysis, phishing detection, and incident response. These applications focus on applying LLMs to distill unstructured text, classification, and agentic automation tasks. Fu et al. \citet{fu2025rasevalcomprehensivebenchmarksecurity} recently introduced RAS-Eval, a comprehensive benchmark to evaluate security-focused LLM agents in real-world environments, which aligns with our goal of realistic evaluation, but focuses on broader security tasks beyond rule generation.

In addition to code quality metrics and datasets, Kapoor et al. \citet{kapoor2024ai} take a cost benefit analysis point of view, that takes into account not only the benefits in increasing ability to fix bugs, but also the cost of such benefits in dollars to run the LLM. 

As with tests passing or failing for code quality, the existing metrics for intrusion detection systems have largely revolved around accuracy in terms of detection rates, true and false positive rates, and false negative rates (which, of course, are very hard to estimate); see, for example, \citep{milenkoski2015evaluating}. In real life, false negative rates are very rarely counted, as they are incredibly hard to track; after all, we do not know what we do not know. In addition, in practice, true positives that do not appear as results for any other detection rule are often valued over rules with similar precision and recall that do not have this property. 

The work builds on these foundational works, while making several novel contributions by introducing a benchmark framework specifically to address the challenges posed by autonomous detection rule generation in the security space. These contributions include:

\begin{itemize}
    \item \textbf{Measuring the Operational Viability of Detection Rules:} Previous work is mainly focused on functional correctness and the generation of secure code. Our framework introduces an end-to-end pipeline to evaluate detection rules holistically. We combine detector validation (e.g. functional correctness) with a real-world corpus for retro-hunting to surface operational noise thresholds, expected maintainability, and, ultimately, performance guardrails, all of which are integrated into a final evaluation score.
    \item \textbf{Built-in Adversarial Robustness:} To better address the brittleness commonly associated with security detection rules, our benchmark incorporates logic to measure brittleness and provide feedback to the code generation agent in an effort to reduce trivial adversarial bypasses.
    \item \textbf{Open-Source Evaluation Framework:} Finally, we plan to open-source the complete benchmark in the hope that it provides a framework for security researchers to test their own rule generation agents (e.g., custom domain-specified language, YARA, etc.). We have provided an adaptable modular pipeline to promote reproducible research and a standardized set of metrics to evaluate code generation agents in the security space.
\end{itemize}

\section {Autonomous Detection Engineer, AD\'E} \label{Sec:ADE}

AD\'E (pronounced ``Ah-Day'') is Sublime Security's LLM-based system to autonomously generate email threat detection rules in MQL. This agentic system is given access to the same tools and much of the same knowledge available to human detection engineers at Sublime Security in order to function as an autonomous detection engineer. AD\'E produces rules to be approved by a reviewer in much the same way that a human would ask a reviewer to review a PR. Unlike a human engineer, AD\'E is restricted to investigating a particular email and creating or modifying a rule inspired by that message. 

AD\'E employs a model-agnostic architecture built around a curated knowledge base containing detection engineering best practices, existing rule patterns, known attacker behaviors and TTPs, and comprehensive MQL documentation. Beyond the knowledge base, AD\'E has access to validation tools, including dynamic file analysis, computer vision, natural language understanding (NLU) capabilities, sender reputation analysis, and comprehensive threat hunting tools. AD\'E can delegate specialized subtasks to focused subagents for activities such as rule critique and hunt result analysis, creating a sophisticated multi-agent workflow. This ``Knowledge, Tools, and Subagents'' architecture reflects a strategic decision to avoid traditional fine-tuning approaches that bind system capabilities to specific model weights. Instead, this design choice allows AD\'E to be used with any foundation model. 

When AD\'E receives a malicious email sample, it follows a systematic investigative process. The system first conducts a comprehensive analysis of the email, examining headers, content, attachments, links, and sender characteristics to identify specific attack vectors and indicators of compromise. It leverages Sublime's analysis tools, including dynamic file analysis for attachments, link analysis for URL examination, and sender profiling to understand the email's context within the recipient's environment. Based on this investigation, AD\'E identifies key TTPs and indicators that could serve as detection points, such as suspicious sender patterns, content obfuscation techniques, or malicious attachment characteristics.

Following this analysis, AD\'E searches Sublime's existing rules for detection rules targeting similar threats or employing comparable indicators. This research phase allows the system to understand existing detection approaches and identify potential coverage gaps. AD\'E then generates candidate MQL detection rules, drawing from its knowledge base of detection patterns while incorporating the specific indicators discovered during its investigation. The system creates rules that target multiple attack vectors simultaneously, such as combining sender reputation checks with content analysis and attachment scanning to maximize detection effectiveness.

The key to AD\'E's approach lies in its iterative refinement capability to address a fundamental challenge in long-context agentic tasks: maintaining focus and achieving convergence without losing track of objectives. Traditional LLM agents struggle with extended workflows because they lack mechanisms to receive corrective feedback when intermediate steps fail \citep{cemri2025multiagentllmsystemsfail}. AD\'E solves this through continuous integration of evaluation framework components as feedback signals. After generating initial detection rules, AD\'E validates syntactical correctness through automated MQL syntax checking, tests rules against the original malicious sample, and initiates threat hunts across historical email data. Different subagents then analyze hunt results to provide detailed feedback on detection accuracy (true positive rates), operational noise (false positive rates), and rule robustness. This creates a closed feedback loop where error messages and evaluation metrics become guidance signals, allowing AD\'E to iteratively refine detection logic, adjust false positive reduction conditions, and incorporate additional indicators until converging on effective rules with acceptable hunt results. This feedback-driven convergence mechanism enables AD\'E to complete complex, multi-step detection engineering tasks that would otherwise fail due to context drift or objective misalignment.
\begin{figure*}[tb]
\centering
\begin{tikzpicture}[scale=0.8, transform shape,
    box/.style={rectangle, rounded corners=2mm, minimum width=2.5cm, minimum height=0.9cm, text centered, draw=black, thick, align=center, font=\footnotesize, fill=white},
    toolbox/.style={rectangle, rounded corners=2mm, minimum width=2.5cm, minimum height=6cm, text centered, draw=black, thick, align=center, fill=gray!5},
    dottedbox/.style={rectangle, rounded corners=1mm, minimum width=2.2cm, minimum height=0.5cm, text centered, draw=black, dotted, font=\tiny, align=center},
    arrow/.style={thick,->,>=stealth},
    looparrow/.style={very thick,->,>=stealth}
]

\node [box] (rulefn) at (2,7) {Rule FN};
\node [box] (analysis) at (2,5.5) {Message Analysis};
\node [box, minimum width=3cm] (proposes) at (2,3.5) {Proposes New or\\Revises Candidate Rule};
\node [diamond, aspect=2, draw] (decision) at (6,2.25) {Good Enough?};

\node [box, minimum width=3cm] (surface) at (2,1) {Surface Final Rule and\\Hunt Performance};
\node [toolbox] (tools) at (-3,3.2) {};
\node at (-3,5.7) {\footnotesize\textbf{Tools}};
\node [dottedbox] at (-3,5) {Check Brittleness};
\node [dottedbox] at (-3,4.25) {Research Existing\\Rules and Coverage};
\node [dottedbox] at (-3,3.5) {Check Syntax};
\node [dottedbox] at (-3,2.75) {Run Rule Against\\Message};
\node [dottedbox] at (-3,2) {Hunt and Analyze\\Results};
\node [dottedbox] at (-3,1.25) {Ask subagents for\\feedback};

\draw [arrow] (rulefn) -- (analysis);
\draw [arrow] (analysis) -- node[right, font=\tiny] {\shortstack{Key Flagging\\Indicators}} (proposes);
\draw [arrow] (proposes) -- (decision);
\draw[arrow] (decision) -- node[above] {\shortstack{No}} (proposes);
\draw[arrow] (decision) -- node[below] {\shortstack{Yes}} (surface);
\draw[line width=1pt, ->, >=stealth, black] 
    (proposes.west) to[out=150, in=30] 
    node[midway, above, font=\scriptsize, black] {Tool Call}
    (tools.east);
\draw[line width=1pt, ->, >=stealth, black] 
    (tools.east) to[out=-30, in=-150] 
    node[midway, below, rotate=90, anchor=east, font=\scriptsize, black] {Tool Response + Feedback}
    (proposes.west);
\end{tikzpicture}
\caption{AD\'E (Autonomous Detection Engineer) Workflow}
\label{fig:ADE_workflow}
\end{figure*}
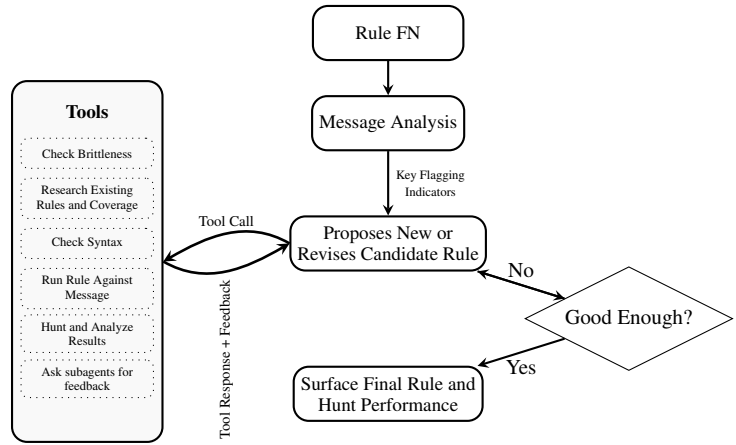

Measuring AD\'E's rule quality is critical for AD\'E's continued development. Without clear metrics defining what constitutes a ``good'' detection rule, and how AD\'E's rules compare to human written rules, it becomes impossible to systematically improve AD\'E's rule generation or validate its effectiveness against human-generated baselines. Indeed, explicitly telling AD\'E what a ``good'' rule looks like is critical to the system; for example, AD\'E has a tool to hunt and analyze the results exactly to work on the true positive vs. false positive quality of its results.

\section{LLM Detection Measurement Framework}
The main goal of this work is to present a framework for measuring the quality of a detection generated by an LLM in comparison to held out detection rules generated by humans. We measure the quality of a detection by measuring the ability of the detection to find malicious emails, the robustness of the query, and the economic efficiency of the query separately. We then use ``holdouts'' of human written detections to compare the quality of AD\'E written detections with a known, high quality baseline. 

It should be noted that the rules written by experts at Sublime Security are meant to flag email messages that could be malicious, but are not typically a trigger for an action based on the rule alone. They instead trigger further triage, by either a model or a human analyst, so while precision is important, it is not the only metric on which we base our measurements, and we do not expect rules to be perfect in precision, even though we strive to reduce false positves. 

In this section we give detailed descriptions of the accuracy metrics that we use to judge rules and intuition about why we chose these metrics. In the results section, we present the results of these metrics on rules held out from AD\'E. 
\subsection{Detection Accuracy}
The foremost concern of any system of security rules, no matter whether generated by humans of AI, is detection accuracy. We measure detection accuracy by measuring the number of true and false positives a rule generates. We are able to measure these counts using an expert human-labeled dataset of emails maintained by Sublime Security for research and development purposes. Once we have ascertained what messages a rule flags and whether those messages are true positives (malicious) or false positives (benign), we also check whether true positives are unique to the detection rule at hand, that is, whether they have been found by any other rules in the core feed, with the idea that malicious emails that are not found any other way are of additional value over ones that are flagged by many rules. In the end, we compute a score of the effectiveness of a rule by 
\begin{equation}
\text{Score} = \frac{1}{2}\left(\frac{\#\mathrm{TP}}{\#\mathrm{TP}+\#\mathrm{FP}} + \frac{\#\mathrm{unique~TP}}{\#\mathrm{TP} + \#\mathrm{FP}}\right)
\label{eq:detection_score_alt}
\end{equation}
This is the average of the standard precision score and the precision of unique true positives, not found by any other rule. It is an attempt to balance the rule's ability to find unique true positives, and thus increase the system's overall recall with the standard classifier effectiveness measurement of precision. In security, we often struggle with our ability to measure false negative rates, as we don't know what we don't know. It is especially hard to understand false negatives in AD\'E's case, as we have not given criteria as to what the rules generated by AD\'E ``should'' catch. Thus, we focus on the more typical, precision-based detection metrics.

\subsection{Economic Cost of Syntactic Correctness}
As mentioned in section \ref{Sec:ADE} above, the pipeline includes an MQL validation step that must be passed before the agent can proceed to the retro-hunt evaluation. Failing to pass this validation step causes the agent to enter a potentially costly retry loop, which impacts both computational and economic resources. The economic implications of LLM-based coding are further explored by Miserendino et al. \citet{miserendino2025swelancerfrontierllmsearn}, who investigate whether frontier LLMs can perform cost-effective "freelance" software engineering tasks, providing additional context for our cost-benefit analysis.

The \textit{pass@}$k$ metric\citep{chen2021evaluatinglargelanguagemodels} has become a standard to measure \textit{functional} correctness for code generated by an LLM agent. However, our use case requires \textit{syntactical} correctness, as a syntactically invalid rule cannot be evaluated. However, each validation failure incurs an economic cost associated with another attempt at generating syntactically correct code.

To account for this economic impact, we adapt the ``cost-to-pass'' framework proposed by Erol et al. \citet{erol2025cost}. The authors propose a metric that calculates the expected monetary cost to obtain a correct solution as defined as:
\begin{equation}
v(m,p) = \frac{C_m(p)}{R_m(p)}
\label{eq:cost_to_pass}
\end{equation}
is the cost per attempt, and $R_m(p)$ is the success rate of that attempt.

We adapt this framework to fit our goal of minimizing the number of turns required to achieve syntactic correctness. Here, we define the success rate $R_m(p)$, the \textit{pass@1} rate of our MQL validator, and the cost per attempt $C_m(p)$ as the cost of generating new MQL logic for that attempt. 

In our testing, we apply this metric to help convey the real-world cost of autonomously generating valid rules. This is calculated as:

\begin{equation}
\text{Total Cost} = C_m(p) \times k
\label{eq:total_cost}
\end{equation}

where $k$ is the number of attempts required for the agent to generate MQL to pass the validation step. By calculating both the \textit{pass@}$k$ rate and the associated cost per generation, we can provide organizations with a metric that conveys the economic impacts of autonomously created security detections.

\subsection{Robustness of Query}
We develop a heuristic to measure the robustness of human-generated and LLM-generated email security rules, inspired by the defensive programming techniques in RobGen\citep{li2025enhancingrobustnessllmgeneratedcode}, but adapted to address the unique challenges of detection engineering. While Li et al. determines robustness by examining the absence of conditional logic and improper error handling in generated code, our approach involves evaluating how susceptible detection rules are to adversarial evasion attempts. Recent work on LLM evaluation in cybersecurity contexts, such as CWEval \citep{peng2025cweval}, has shown that measuring code robustness, encompassing both functionality and security (e.g. resistance to vulnerabilities), is crucial to mitigating operational risks when deploying LLM-generated in production environments. In \citep{peng2025cweval} and \citep{fu2024constrained}, the researchers introduce metrics to evaluate both the functionality and the security of the generated code.

These guiding principles apply to our use case (e.g., rule-based security detections), where defenders are in an active cat-and-mouse game with attackers who are actively attempting to evade these detectors through low-effort bypasses.

We assume that direct string matches on fields such as IP addresses, domains, or hash values indicate brittleness, as these patterns may be bypassed through trivial evasions by attackers. Conversely, we view behavioral indicators (sender prevalence, domain reputation scores, natural language understanding) and fuzzy matching as signs of robustness, as they present generalizable solutions to variations in attack patterns.

To quantify this, we compute a brittleness score $B \in [0, 100]$ that derives the score from the ratio of rewards (for robust patterns) to penalties (for brittle patterns). This ratio is mapped to a 0-100 scale using a logistic function. The formula is:

\begin{equation}
B = \frac{100}{1 + e^{k\left(\frac{R}{P} - x_0\right)}}
\label{eq:brittleness}
\end{equation}

This metric accounts for the challenge all rule-based detections face in balancing brittle logic (\textit{to avoid FPs}) and robust logic (\textit{to avoid adversarial evasions}).

The robustness score is then:
\begin{equation}
\text{Robustness} = \frac{100 - B}{100} = 1 - \frac{B}{100}
\label{eq:robustness}
\end{equation}
Our approach differs from previous efforts to define code robustness metrics that focus on input validation and exception handling, because as DefenderBench demonstrates, the security domain requires robustness that accounts for resilience against adversarial evasions rather than accidental errors.\citep{zhang2025defenderbench}

\begin{table*}[!htbp]
\centering
\caption{List of Detectors Tested}
\label{tab:rules}
\footnotesize
\setlength{\tabcolsep}{4pt}
\begin{tabular}{|p{4cm}|p{2.5cm}|p{4cm}|p{2.5cm}|}
\toprule
\textbf{Rule Name} & \textbf{Attack Type} & \textbf{Detection Method} & \textbf{TTPs} \\
\midrule
AFF from freemail/sus TLD & BEC/Fraud & Header, Content, Sender & Social eng. \\
Callback Phishing via PDF & Callback Scam & File Analysis, Exif Analysis & OOB pivot \\
Embedded JS in SVG & Malware & File, Sender Analysis & Scripting \\
EML w/ JS in SVG & Phish/Malware & File, Sender Analysis & Scripting/Evasion \\
HTML smuggling w/ atob & Malware & HTML analysis & HTML smuggling \\
BEC w/ Reply-to mismatch & BEC/Fraud & Header, Content, Sender & Evasion \\
Coinbase impersonation & Cred. Phishing & Header, Content, Sender & Brand Imp \\
DocuSign image lure & Cred. Phishing & Computer Vision, Content & Brand Imp \\
Extortion/sextortion & Extortion & Content, Header Analysis & Social eng. \\
Fake voicemail (untrusted) & Cred. Phishing & Content, URL Analysis & Social eng. \\
HTML smuggling & Cred. Phishing & HTML analysis & HTML smuggling \\
MS Dynamics 365 phishing & Cred. Phishing & File Analysis, OCR & Evasion \\
Open Redirect: Google /url & Cred. Phishing & Header, File Analysis & Evasion \\
QR Code w/ sus indicators & Cred. Phishing & Computer Vision & QR/Social eng. \\
Scam: Piano Giveaway & BEC/Fraud & Content Analysis & Social eng. \\
Spam: Fake photo share & Spam & Content, Sender Analysis & Social eng. \\
Explicit Google Group & Spam & Content, Sender Analysis & Social eng. \\
Suspected Lookalike domain & BEC/Fraud & Content, Sender Analysis & Evasion \\
\bottomrule
\end{tabular}
\end{table*}

\section{Experimental Setup}
Our goal is to illustrate and exercise the measurement of LLM-generated cybersecurity rule quality using held-out rules, for which we use AD\'E and the corpus of emails shared with Sublime Security as an illustrative example.
\subsection{Scenario}
We use AD\'E, described in Section \ref{Sec:ADE} as the tool to illustrate our evaluation framework. The primary scenario is a holdout validation process designed to compare AD\'E to a human baseline. We have curated a set of human written rules, given in table \ref{tab:rules} designed to cover a broad range of tactics, techniques, and procedures (TTPs) from Sublime Security's production core feed \citep{corefeed}. We then presented AD\'E with a TP, also human curated to be representative, that would have been caught by the held-out human written rule. We then measure the effectiveness of the AD\'E generated rule in comparison to the human written held-out rule.
\subsection{Dataset}
We are able to test AD\'E's performance by comparing to held out rules from the Sublime Security Core Feed, removing rules that AD\'E knows about one at a time and asking it to develop rules for examples of messages that would flag with the held out rule. 

We chose a static list of rules to hold out in all runs of our test, with the goal of choosing rules that are likely to have been flagged with messages in our dataset and which represent a variety of TTPs and attacker goals. The rules chosen are listed in Table \ref{tab:rules}.

Our test environment consists of approximately $45,000$ email messages that have been submitted to Sublime Security for analysis over a 14 day period. For this purpose, we extended our labels by hand labeling the relevant emails as malicious or benign, augmenting our preexisting labeled data set with additional relevant labels. 
\section{Results} \label{sec:Results}
Here we examine the results of our measurement framework comparing rules from the Sublime Security core feed and those created by AD\'E in response to a single instance of the core feed rule, as chosen by a human expert to be representative. Of course, this evaluation is somewhat apples to oranges, since we tell AD\'E to address a single malicious email, but the expert humans who wrote the core rules have the goal of addressing a broad behavior based on a large collection of data. The complete evaluation results are given in tables \ref{tab:human}, \ref{tab:ade}, and \ref{tab:comparison}.
\begin{table}[htbp]
\centering
\caption{Performance Metrics for Human-Generated Rules}
\label{tab:human}
\resizebox{\textwidth}{!}{%
\begin{tabular}{|p{5.5cm}|c|c|c|c|c|}
\toprule
\textbf{Name} & \textbf{Hits} & \textbf{TPs} & \textbf{FPs} & \textbf{Unique TPs} & \textbf{Score} \\
\midrule
Callback solicitation via pdf file & 765 & 756 & 9 & 747 & 0.982 \\
EML w/ Javascript in SVG File & 276 & 276 & 0 & 274 & 0.996 \\
HTML smuggling with atob & 16 & 14 & 2 & 12 & 0.813 \\
BEC w/ Reply-to mismatch & 1558 & 1557 & 1 & 793 & 0.754 \\
Brand impersonation: Coinbase & 35 & 33 & 2 & 31 & 0.914 \\
Fake voicemail notification & 344 & 304 & 40 & 264 & 0.826 \\
MSFT Dynamics 365 form phish & 57 & 57 & 0 & 24 & 0.711 \\
Scam: Piano Giveaway & 100 & 99 & 1 & 57 & 0.780 \\
Spam: Fake photo share & 240 & 231 & 9 & 208 & 0.913 \\
Explicit Google Group Invite & 10 & 10 & 0 & 10 & 1.000 \\
Suspected Lookalike domain & 10 & 8 & 2 & 8 & 0.800 \\
\bottomrule
\end{tabular}%
}
\end{table}

\begin{table}[htbp]
\centering
\caption{Performance Metrics for AD\'E-Generated Rules}
\label{tab:ade}
\resizebox{\textwidth}{!}{%
\begin{tabular}{|p{5.5cm}|c|c|c|c|c|}
\toprule
\textbf{Name} & \textbf{Hits} & \textbf{TPs} & \textbf{FPs} & \textbf{Unique TPs} & \textbf{Score} \\
\midrule
Callback solicitation via pdf file & 31 & 31 & 0 & 23 & 0.871 \\
EML w/ Javascript in SVG File & 239 & 239 & 0 & 238 & 0.998 \\
HTML smuggling with atob & 4 & 4 & 0 & 1 & 0.625 \\
BEC w/ Reply-to mismatch & 35 & 35 & 0 & 4 & 0.557 \\
Brand impersonation: Coinbase & 132 & 116 & 16 & 75 & 0.723 \\
Fake voicemail notification & 6 & 6 & 0 & 5 & 0.917 \\
MSFT Dynamics 365 form phish & 56 & 56 & 0 & 24 & 0.714 \\
Scam: Piano Giveaway & 38 & 38 & 0 & 18 & 0.737 \\
Spam: Fake photo share & 202 & 202 & 0 & 187 & 0.963 \\
Explicit Google Group Invite & 7 & 7 & 0 & 7 & 1.000 \\
Suspected Lookalike domain & 6 & 6 & 0 & 6 & 1.000 \\
\bottomrule
\end{tabular}%
}
\end{table}

\begin{table}[htbp]
\centering
\caption{Brittleness and Cost Comparison}
\label{tab:comparison}
\resizebox{\textwidth}{!}{%
\begin{tabular}{|p{5.5cm}|c|c|c|c|}
\toprule
\textbf{Rule Name} & \textbf{Brittleness (AD\'E)} & \textbf{Brittleness (Human)} & \textbf{Cost (AD\'E \$)} & \textbf{AD\'E pass@k} \\
\midrule
Callback solicitation via pdf file & 69.8 & 68.3 & 3.11 & 2 \\
EML w/ Javascript in SVG File & 68.2 & 72.3 & 5.13 & 3 \\
HTML smuggling with atob & 68.0 & 69.9 & 3.48 & 3 \\
BEC w/ Reply-to mismatch & 66.6 & 66.4 & 2.07 & 2 \\
Brand impersonation: Coinbase & 71.7 & 59.8 & 2.54 & 2 \\
Fake voicemail notification & 68.9 & 71.7 & 2.55 & 2 \\
MSFT Dynamics 365 form phish & 67.9 & 66.3 & 1.51 & 1 \\
Scam: Piano Giveaway & 70.6 & 72.5 & 2.07 & 2 \\
Spam: Fake photo share & 66.5 & 70.6 & 2.41 & 2 \\
Explicit Google Group Invite & 70.6 & 70.1 & 1.65 & 1 \\
Suspected Lookalike domain & 71.2 & 69.1 & 1.51 & 1 \\
\bottomrule
\end{tabular}%
}
\end{table}

As one might expect from the constraints of AD\'E's design, the rules that it finds are much more narrowly tailored, finding both fewer overall messages and fewer known TPs, but also many fewer known FPs. This makes sense in that AD\'E is given a single message to work from, rather than the corpus of messages and idea about a set of behaviors to catch that humans use. Thus, AD\'E has a narrower view of the universe of malicious messages one might try to catch with this rule than the human who wrote the human rule. The fundamental security problem that ``you don't know what you don't know'' very much applies to AD\'E as well, since it gets only some of the problem set for humans to solve. 

\begin{figure}[htbp]
\centering
\includegraphics[width=0.75\linewidth]{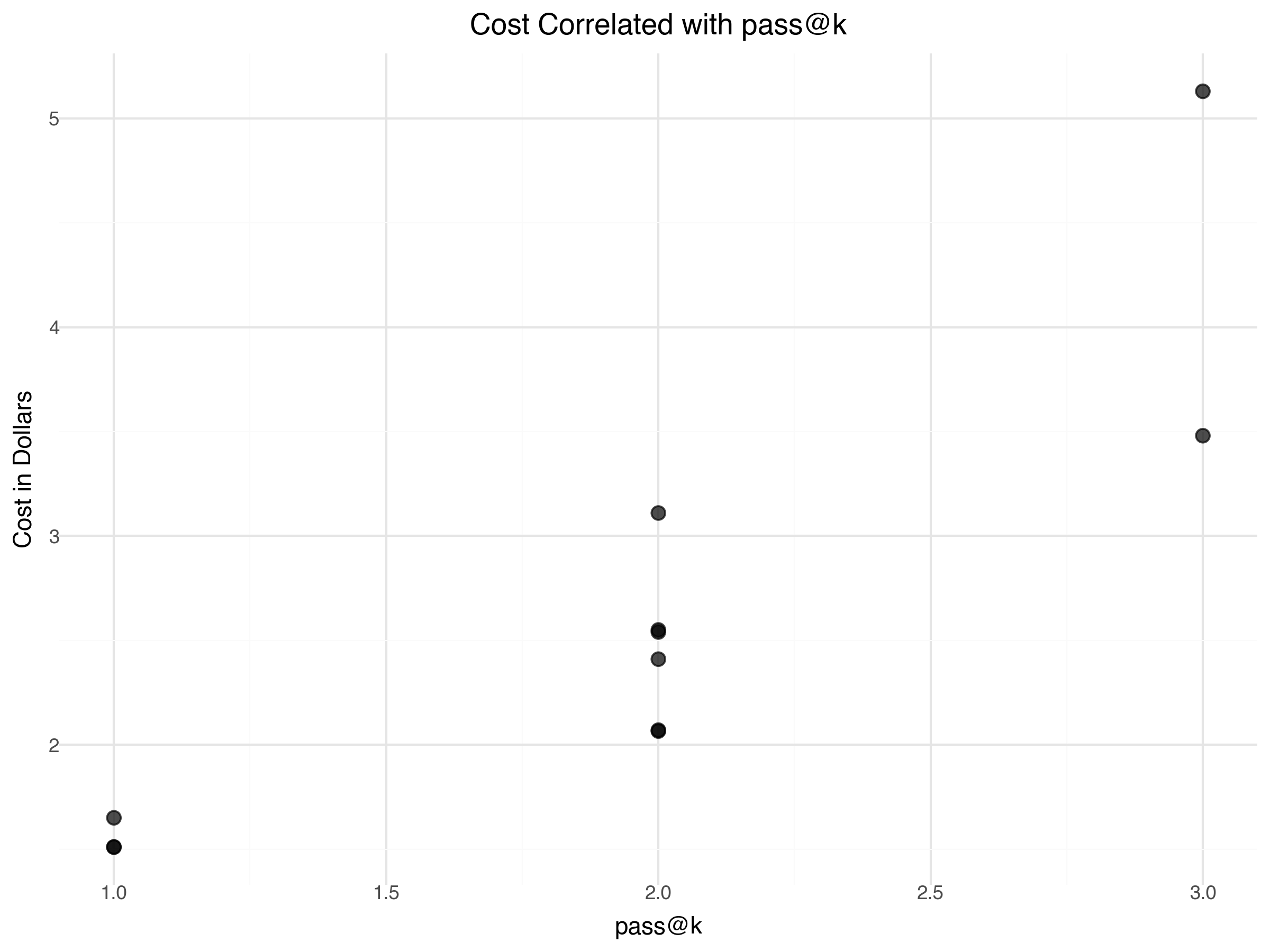}
\caption{Cost of running the LLM as a function of \textit{pass@}$k$.}
\label{fig:pass_at_k}
\end{figure}

One result to note is that the pass@k metric correlates quite closely with the actual dollar cost of running the LLM to produce the result, as expected, see Figure \ref{fig:pass_at_k}. 

AD\'E and humans are quite comparable in the brittleness of the rules they write. Security rules are, by their very nature, somewhat brittle, with difficult trade-offs about TPs, FPs, and FNs, suggesting that the base LLM is able to combine the information and tools it is given about query quality to produce high-quality queries.

One additional interesting discovery that we made as we compared AD\'E's rules with the human written rules is that AD\'E is much more inclined to write a lot of helpful comments than most human authors. We provide in Figure \ref{fig:rule_comparison} a side-by-side view of a human written rule and the rule AD\'E wrote to illustrate the differences described here both in the breath of rules and the comments provided\begin{figure*}[htbp]
\centering
\caption{Comparison of AD\'E-generated and Human-generated rules for detecting HTML smuggling}
\label{fig:rule_comparison}
\begin{minipage}[t]{0.48\textwidth}
\centering
\textbf{AD\'E Generated Rule}
\lstset{basicstyle=\scriptsize\ttfamily, numbers=left, numberstyle=\tiny}
\begin{lstlisting}[frame=single]
type.inbound

// SECTION 1: EML attachment with SVG content
and any(attachments,
  .content_type == "message/rfc822"
  and strings.icontains(file.parse_text(.).text, ".svg")
)

// SECTION 2: Base64 encoded JavaScript 
// obfuscation patterns
and any(attachments,
  .content_type == "message/rfc822"
  and any(beta.scan_base64(file.parse_text(.).text),
    (
      strings.ilike(.,
        "*atob*",
        "*eval*",
        "*fromCharCode*"
      )
      or strings.ilike(.,
        "*window.location*",
        "*document.location*"
      )
      or regex.icontains(.,
        "parseInt.*charCodeAt")
    )
  )
)

// SECTION 3: Recipient targeting indicator
and any(attachments,
  .content_type == "message/rfc822"
  and any(recipients.to,
    strings.icontains(file.parse_text(..).text,
      .email.email
    )
    and .email.domain.valid
  )
)

// SECTION 4: Sender validation
and (
  profile.by_sender().prevalence in (
    "new",
    "outlier"
  )
  or not profile.by_sender().solicited
  or not headers.auth_summary.dmarc.pass
)
\end{lstlisting}
\end{minipage}
\hfill
\begin{minipage}[t]{0.48\textwidth}
\centering
\textbf{Human-generated Rule}
\lstset{basicstyle=\scriptsize\ttfamily, numbers=left, numberstyle=\tiny}
\begin{lstlisting}[frame=single]
type.inbound
and any(attachments,
  (.content_type == "message/rfc822" 
   or .file_extension =~ "eml")
  and (
    any(file.parse_eml(.).attachments,
      .file_extension in~ ("svg", "svgz")
      and (
        (
          strings.ilike(file.parse_text(.).text,
            "*onload*",
            "*window.location.href*",
            "*onerror*",
            "*CDATA*",
            "*<script*",
            "*</script*",
            "*atob*",
            '*location.assign*',
            '*decodeURIComponent*'
           )
          or regex.icontains(file.parse_text(.).text,
            '<iframe[^\>]+src\s*=\s*\"data:[^\;]+;base64,'
          )
          or any(beta.scan_base64(file.parse_text(.).text),
            strings.ilike(.,
              "*onload*",
              "*window.location.href*",
              "*onerror*",
              "*CDATA*",
              "*<script*",
              "*</script*",
              "*atob*",
              '*location.assign*',
              '*decodeURIComponent*'
            )
           )
        )
        or // Additional sender logic...
      )
    )
  )
)
\end{lstlisting}
\end{minipage}
\end{figure*}

\section{Conclusions and Future Work}
There are two main branches of future work that we intend to pursue with the framework for evaluating security rules given here: further developing the framework and using the framework to further develop and refine AD\'E. 

The hunt results of true and false positives reveal that AD\'E tends to find fewer results than the human-generated held-out rules, even though it makes quite robust rules and is reasonably economically efficient. This is not surprising, given the design of AD\'E to work on a single malicious email at a time. We can use the measurement framework described here for iterative improvements to AD\'E, for example finding an effective method to broaden AD\'E's scope to build detections that rival the recall of human written detections. 

In addition, we would like to expand the testing framework described here in a number of ways. In real life, email security systems are expected to filter spam and graymail (potentially unwanted bulk mailing list messages) from mailboxes, in addition to filtering malicious messages. It would be useful to expand this framework to other categories that are of interest with a multimodal version of the framework. Furthermore, it would be helpful to include adversarial robustness tests in addition to the current evaluations of the rules, where we evaluate the generated rules against a number of automutated threat variants in addition to looking for static signs of adversarial robustness. Adversarial robustness tests likely would further help us improve AD\'E in making broader rules in the way that humans currently do.

\end{document}